\input epsf.tex
\documentstyle[aaspp4]{article}
\def\etal{{\it et al. }}
\def\spose#1{\hbox to 0pt{#1\hss}}
\def\lsim{\mathrel{\spose{\lower 3.0pt\hbox{$\mathchar"218$}}
     \raise 2.0pt\hbox{$\mathchar"13C$}}}
\def\gsim{\mathrel{\spose{\lower 3.0pt\hbox{$\mathchar"218$}}
     \raise 2.0pt\hbox{$\mathchar"13E$}}}
\def\msun{{\rm\,M_\odot}}

\begin{document}
\title{Evolution of Debris of a Tidally Disrupted Star by a Massive Black Hole:
Development of a Hybrid Scheme of the SPH and TVD Methods}
\author{Hyung Mok Lee}
\affil{Department of Earth Sciences, Pusan National University, Pusan 609-735,
Korea}
\and
\author{Sungsoo S. Kim}
\affil{Institute for Basic Sciences, Pusan National University, Pusan 609-735,
Korea}

\begin{abstract}
The evolution of the stellar debris after tidal disruption due to the super
massive black hole's tidal force is difficult to solve
numerically because of the large dynamical range of the problem.
We developed an SPH (Smoothed Particle Hydrodynamics) - TVD (Total
Variation Diminishing) hybrid code in which the SPH is used to cover a widely
spread debris and the TVD is used to compute the stream collision more
accurately.  While the code in the present form is not sufficient to
obtain desired resoultion, it could provide a useful tool in studying
the aftermath of the stellar disruption by a massive black hole.
\end{abstract}

\section{INTRODUCTION}

The stars in the vicinity of the Super Massive Black Hole (SMBH), which is
believed to exist at the center of galaxies, are disrupted by the
SMBH's strong tidal force at a rate of one per $10^{3 \sim 4}$ yrs (e.g.
Gooman \& Lee 1989).  It is very important to know the detailed evolution
of the stellar debris because the amount of energy released as a result of
stellar disruption depends on the way of evolution of the debris.

However, the numerical approach to this problem is very difficult.
The debris forms a narrow, long stream in a wide area around the black
hole.  The incoming stream with short orbital periods collides
with the outgoing stream because the orbits around a massive black hole
do not form a closed ellipse.
The precise evolution of the debris depends on the outcome of the
stream collision which requires very high resolution studies.
No single numerical scheme developed so far is able to handle this problem.

Cannizzo, Lee, \& Goodman (1990) argued that the initial, extremely eccentric
orbit of the stellar debris will form a circular disk after experiencing
strong shocks which thermalize the orbital energy and studied the subsequent
evolution of the accretion disk using a time-dependent $\alpha$-disk model.
Kochanek (1994) studied the long-term evolution of the debris stream and
predicted the geometry and physical properties when the stream collides
with itself.  However, the circularization argument by Cannizzo \etal
(1990) and the results of Kochanek's (1994) analytical method remain to be
numerically verified.

Numerical simulations for the evolution of the debris stream were done
by Monaghan \& Lee (1994) and Lee, Kang, \& Ryu (1996).  Although Monaghan
\& Lee (1994) used the SPH (Smoothed Particle Hydrodynamics) method to
follow the evolution of the stellar debris in a huge space, the poor
resolution due to a limited number of particles made a detailed study
for the collision area impossible.  On the other hand, Lee \etal (1996)
concentrated on the hydrodynamics of the collision area of the debris
streams using the TVD (Total Variation Diminishing) scheme.  They concluded
that the stream-stream collisions are expected to circularize the stellar
debris efficiently unless the ratio of the cross sections of the two streams is
much larger than 10.  However, the region of their simulation was restricted
only to the collision area, and the gravitational force by the
black hole was not included in the calculation because the entire volume of
their simulation was very small.

There are usually two approaches to solve the hydrodynamics numerically:
Lagrangian scheme and Eulerian scheme.  The Lagrangian scheme is adequate
for the problems with time-varing extent while Eulerian scheme is better
fitted to the problem of limited extent.  Lagrangian method is suitable to
the overall spread of the debris but Eulerian scheme is a better choice
for the description of stream collision.  Thus we developed a hybrid scheme
in the present paper to study the long term evolution of stellar
debris.  We have adopted Smoothed Particle Hydrodynamics (SPH) as a Lagrangian
scheme and Total Variation Diminishing (TVD) scheme as an Eulerian scheme.

The SPH method was invented to simulate complex hydrodynamic phenomena in
astrophysics (for the details of SPH method, see Monaghan [1992] and references
therein).  It is a particle method in which physical quantaties are described
with non-fixed particles.  Spatial derivatives are calculated by analytical
differentiation of interpolation formulae.  Since the SPH is a particle method,
it can handle spatially complex, wide-area physics more easily than
finite-difference methods which are more accurate for some problems requiring
high resolution.

The TVD method was originally developed by Harten (1983).  It is an explicit,
second-order, Eulerian finite-difference method which solves a hyperbolic
system of the conservation equations.  The key merit of this method is to
achieve the high resolution of a second-order accuracy while preserving
the robustness of the nonoscillatory first-order method.

This paper is organized as follows.
Preliminary simulation using the SPH is first shown in \S \ref{sec:sph},
and the SPH-TVD hybrid procedure used in this study is
described in \S \ref{sec:code}.  The simulation results performed with such
a code are presented in \S \ref{sec:simulation}. The discussions
will be given in \S \ref{sec:discussion}.

\section{SOME RESULTS FROM THE SPH SIMULATIONS}
\label{sec:sph}

\begin{figure}
\plotone{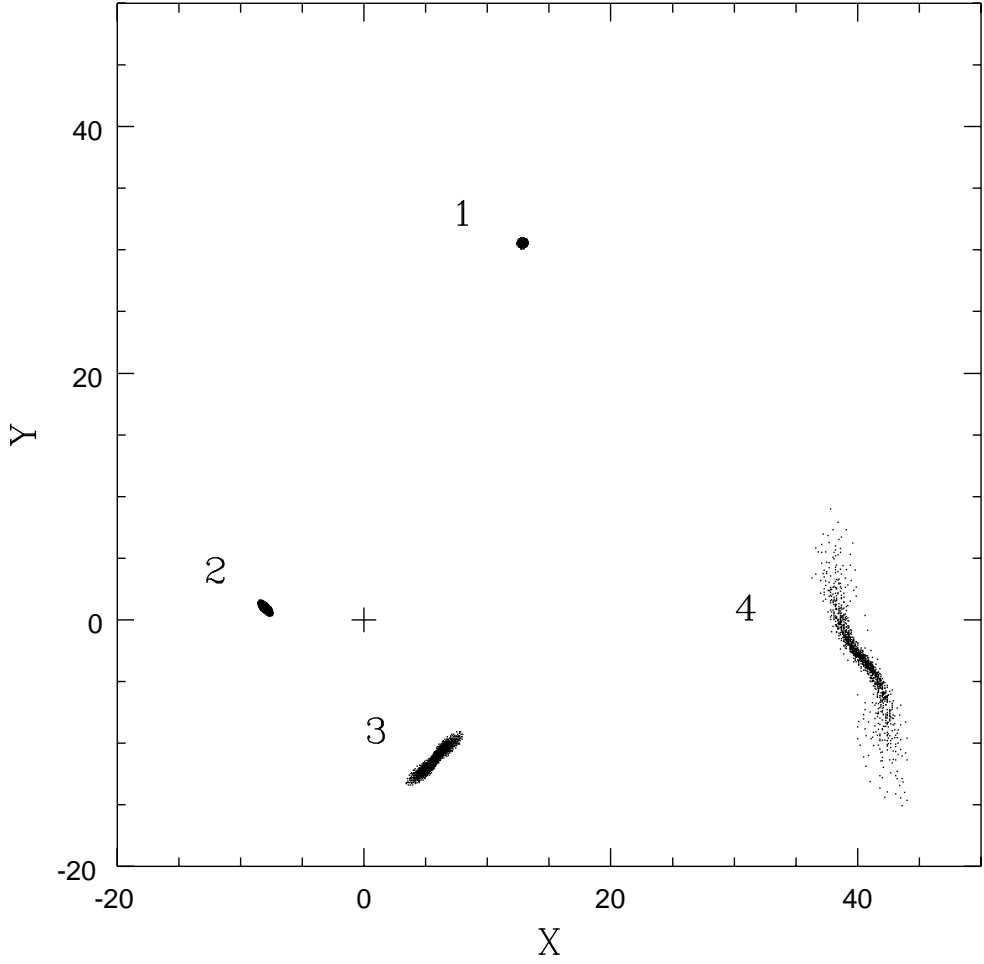}
\caption{\label{fig:sph1}Four snapshots of initial SPH simulation for
the encounter between an SMBH and a normal star.  The location of the SMBH
is indicated with a + sign, and the normal star is being disrupted by the strong
tidal force of the SMBH.  The sizes of the stars are increased by a factor of
10 and the length is in $1.48 \times 10^{12}$ cm.
See the text for simulation parameters.}
\end{figure}

\begin{figure}
\plotone{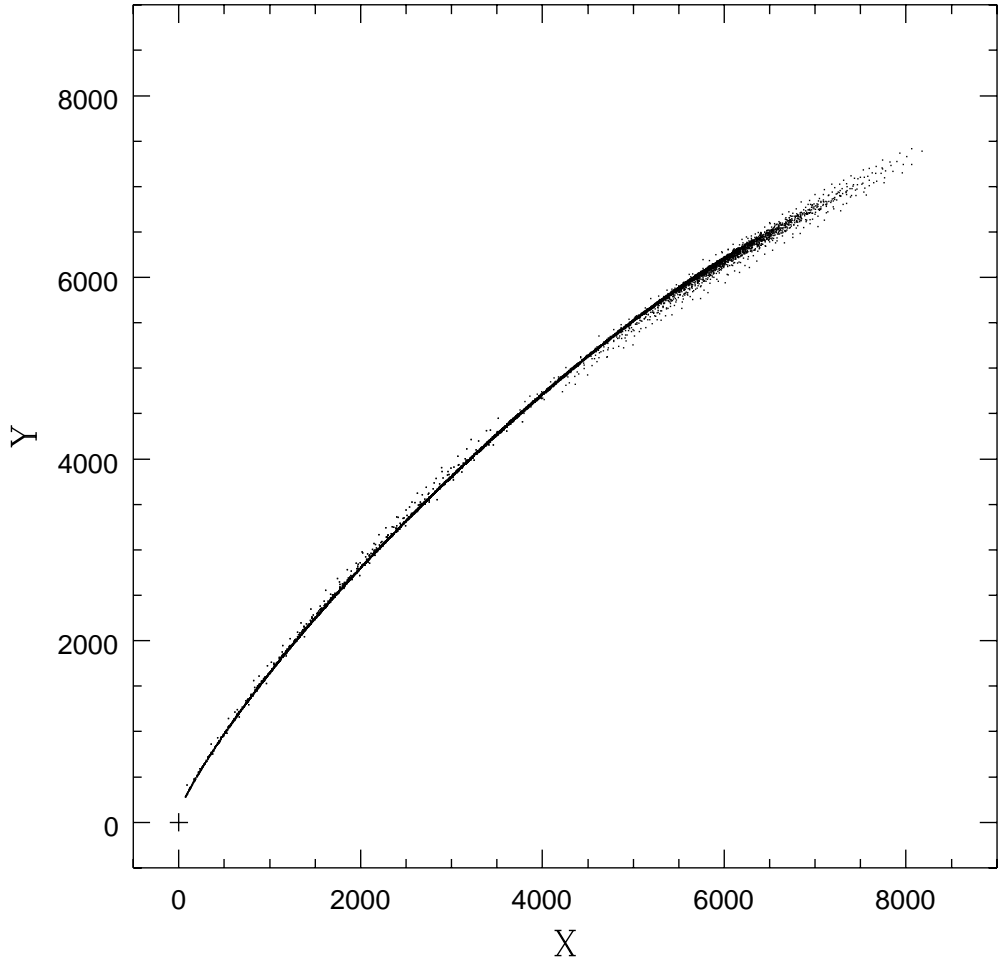}
\caption{\label{fig:sph2}Stream of the stellar debris disrupted by the SMBH.
The disrupted normal star in Figure 1 now forms a long, thin stream as it
passes the apogee, and is approaching near the SMBH for the second time.
The length unit is the same as in Figure 1.}
\end{figure}

\begin{figure}
\plotone{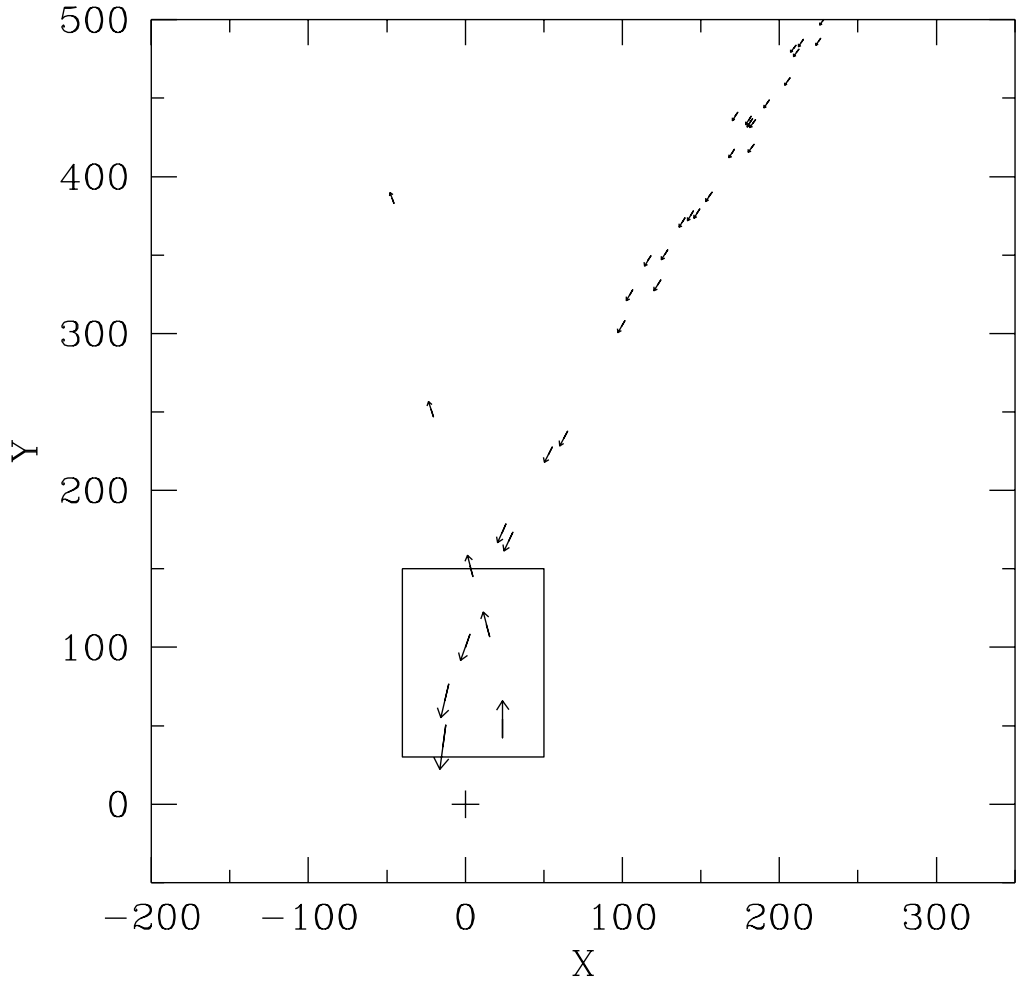}
\caption{\label{fig:sph3}A leading part of the stream of the stellar debris
disrupted by the SMBH.  The head, which already passes by the SMBH for the
second time, collides with its tail near (10,140).  The number of SPH
particles at the collision area is abviously to small for a precise
hydrodynamic calculation.  We put a TVD box (represented by the inner box)
near the collision region to perform a precise hydrodynamic
calculation for the collision with the TVD scheme, whose accuracy is
better for this kind of situation than that of the SPH method.  The length
unit is the same as in Figure 1.}
\end{figure}

A series of snapshots of an SPH simulation for the typical first encounter
between a normal star and an SMBH is shown in Figure \ref{fig:sph1}.
The normal star here is modelled by a polytrope of index 3/2.  The mass
of the SMBH, $M_{\rm BH}$, is $10^7 \msun$, mass of the normal star is
$1 \msun$, and the stellar orbit is chosen as a hyperbola with pericentral
distance of $1.5 \times 10^{13}$ cm with $v_{\infty}=10$ km/s.
The tidal force by the SMBH is so strong that the normal star is completely
disrupted and a long stream of the stellar debris is forming.  This stream is
even more stretched out as it returns to the SMBH for the second time as in
Figure \ref{fig:sph2}.  The head of the stream is composed of the particles
facing the SMBH when the star passes by the SMBH, and thus has smaller
semi-major axes and periods than the tail.  After the head passes by the SMBH
for the second time, it collides with incoming stream as shown in Figure
\ref{fig:sph3}.  However, the number of the SPH particles in the collision
area in our simulation is too small for a precise hydrodynamic description.
Our SPH simulation was done with 9185 SPH particles, and with a current CPU
ability, it is not practicable to perform an SPH simulation with more than
few tens of thousands particles.  This shows the reason why the SPH alone
is not able to handle this specific problem.

It is inevitable to use the SPH method at the beginning of the above
calculation because the debris of the disrupted star is spread over a wide area,
but a supplementary scheme is neccessary to obtain a higher resolution in the
collision area.  For this reason, we set a TVD grid where hydrodynamically
complex phenomena take place as in Figure \ref{fig:sph3}, and conduct
the calculation of the evolution of the debris using the TVD when an SPH
particle comes into the TVD grid.  The SPH is applied again for the material
out of the TVD grid.

\section{THE HYBRID CODE}
\label{sec:code}

Making a hybrid code out of conceptually different two numerical
methods is not a trivial task.  Physical quantities on the SPH-TVD
border should be converted carefully from SPH to TVD and TVD to SPH,
conserving energies and other physical quantities.  Although it is quite
straightforward to convert from the SPH to the TVD, the reverse procedure
requires a bit of numerical consideration to keep
the number of created particles small enough for the whole run.

When the SPH particles approach pericenter for the second time,
due to a limited number of the SPH particles, the resolution length
of those particles is so large that the incident flux to the
TVD box fluctuates unrealistically.  Therefore, we take time-averages
for the flux of incoming SPH particles and artificially create incoming
particles with a constant flux at the SPH-TVD border.  Since the period of
our whole run is short compared to the orbital period of the very tail 
of the stream, the incoming flux may be assumed to be a constant.  (The
actual incoming flux varies like $t^{-5/3}$; see, for example, Rees [1988].)
Then the conversion from SPH to TVD is done with such artificial
SPH particles, and we will call this procedure SPH2TVD.  Detailed procedure
is discussed in \S \ref{sec:SPH2TVD}.

Now the quantities mapped onto the TVD grid are calculated with the TVD code.
For the TVD part, we adopted the code written by Lee \etal (1996).
A modification to the original code is explained in \S \ref{sec:SPH2TVD}.

The time step for the TVD calculation is determined by the velocity of the
fluid in the TVD box.
Although the speed of the stream reaches a maximum at the closest point
to the SMBH, hydrodynamic effects near the SMBH may be neglected because
the pressure force is much smaller than the tidal force and gravitational
force.  Thus by locating the SMBH outside the TVD box and calculating the
motion of the fluid as a collection of free-streaming-particles (particles
with no hydrodynamics), one could save large amount of CPU time and memory.
For this reason, we put the TVD box that encompasses the collision area only
(apart from the SMBH), and at the TVD border facing the SMBH, we convert the
TVD quantities into SPH particles without considering hydrodynamics.
We will call this procedure TVD2SPH, and the details of this procedure
are given in \S \ref{sec:TVD2SPH}.

After the SPH particles that are created at the TVD border facing the SMBH,
they revolve around the SMBH and enter the TVD box for the second time.
This step is identical to the SPH2TVD procedure above.

Finally, the stream collides with its own tail in the TVD box, where
the hydrodynamics is calculated with a higher resolution using TVD method.

It was found that the sequence of the above procedures is important in
determining the accuracy and consistancy of the whole simulation.
The sequence which gives the best result would not be only one, but we find
that the total energy is conserved the best when we make the code with the
following sequence:

Step 1. TVD2SPH : Conversion of physical quantities from TVD grids to SPH
particles.

Step 2. TVD : TVD calculation.

Step 3. SPH : SPH calculation.

Step 4. SPH2TVD : Conversion from SPH particles to TVD grids.

Step 5. SPH stacking : Keeping the number of SPH particles small.

Step 6. Output : Making various outputs.

Step 7. Time step : Determining the time step.

Step 8. Go to Step 1.

In Step 7, the time step of the code is usually determined by the TVD part
since it requires a smaller time step than other procedures.  Then
the SPH quantities are advanced by one time step by integrating the
equation of motion over one time step.  Steps 1 and 4 will be
discussed below in detail.

\subsection{Conversion from SPH to TVD}
\label{sec:SPH2TVD}

\begin{figure}
\plotone{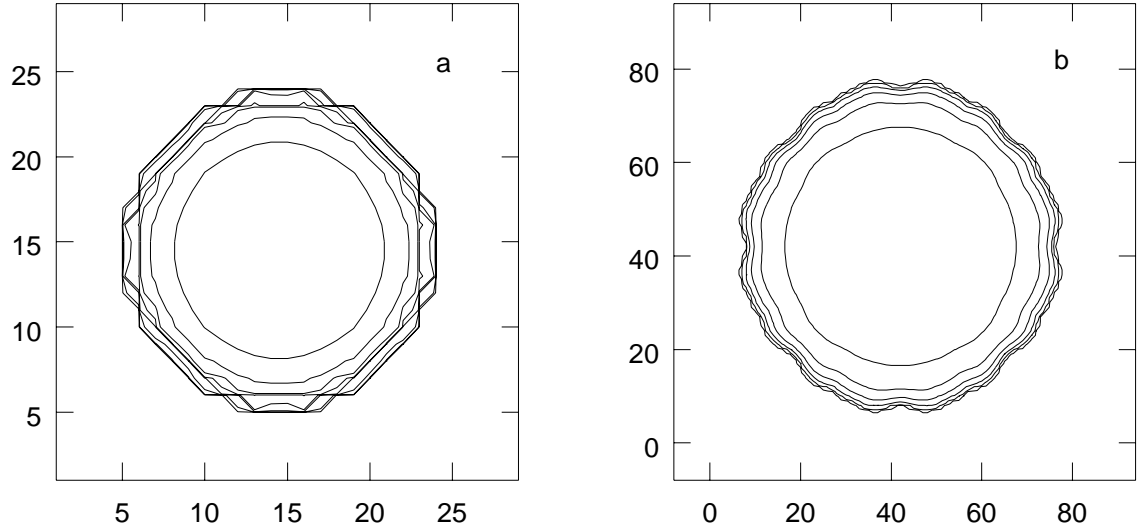}
\caption{\label{fig:poly}Density contour maps generated by mapping a normal star
realized with a polytropic index of 3/2 with 9185 SPH particles
into the TVD grids using interpolation kernel with (a) grid size = $h$,
and (b) grid size = $h/4$, where $h$ is the resolution length of the
interpolation kernel.  To nicely map the SPH particles into the TVD grids
down to the region with a very low density, the grid size should be less
than the resolution length of the SPH interpolation kernel.}
\end{figure}

The initial phase should be calculated by using the SPH where fluid element is 
represented by a collection of particles. We need to convert the particle 
variables into field variables at grid points in order to perform TVD
calculations where high spatial resolution is required.  The
conversion is automatically carried out by the SPH part in the following
manner:
\begin{equation}
\begin{array}{rcl}
	\rho (x,y,z) & = & \sum_i m_i W(r,h)\\
	\rho (x,y,z) {\bf v}(x,y,z) & = & \sum_i m_i {\bf v}_i W(r,h)\\
	u(x,y,z) & = & \sum_i m_i E_{th,i} W(r,h),
\end{array}
\end{equation}
where $\rho$ is the density, $m_i$ is the mass of the i-th particle, $u$ is
the thermal energy per unit volume, $E_{th}$ is the thermal energy per unit
mass of the particle, and $W(r,h)$ is the spherical interpolation kernel with
resolution length $h$. As for $W(r,h)$, we have adopted the spline based
kernel in our code.  The only remaining problem is how to choose the
grid size.  In order to determine the optimal grid size we made a few
experiments.  In Figure \ref{fig:poly}a, we have shown the density contour
of an unperturbed main-sequence star of polytrope of $n=3/2$ realized by
particles using SPH and mapped onto cubical grids of the size equivalent to the
resolution length.  It is clear that the grid size is too large to represent
the low density region in the outer parts of the star.  Thus we need to choose
the grid size smaller than the resolution length $h$ in order to have a
accurate and realistic density distribution.  Increasing the number of
grid points by a factor of 4 on each side gave an adequate representation
as shown in Figure \ref{fig:poly}b.  Repeated experiments showed that the grid
size should be at least four times smaller than the SPH's resolution length.

Since the original TVD program that we adopted did not have the external
gravitational field, we have modified the equation of motion to take into
account the relativistic gravitational field as follows.

Step 1. Compute $\phi^n$ from the velocity field $v_i^n$ at the grid
points.  The velocity comes in because of the relativistic correction,
for which we adopted the following post-Newtonian formula
\begin{equation}
	\phi = {\rm G} M_{\rm BH} \left (-{1 \over r}+{1 \over 2 r^2}
		  +{3 v^2 \over 2 r}+{3 v^4 \over 8} \right ).
\end{equation}

Step 2. Estimate the new velocity that takes into account the effect
of external gravitational field: i.e., 
\begin{equation}
	v_i^n = {\tilde v_i^n} + {1\over 2} {\partial \phi^n\over \partial x_i},
\end{equation}
where ${\tilde v_i^n}$ is the velocity field in the absence of external gravity.

Step 3. Make corrections of hydrodynamical force computed by the TVD:
\begin{equation}
\begin{array}{rcl}
	\rho^{n+1} & = & \rho^n+\Delta\rho^{TVD}(\rho^n,{\tilde v_i^n},
			{\tilde P^n})\\
	v_i^{n+1} & = & v_i^n+\Delta\rho^{TVD}(\rho_i^n,{\tilde v_i^n},
			{\tilde P^n})\\
	P^{n+1} & = & P^n+\Delta P^{TVD}(\rho^n,{\tilde v_i^n},
			{\tilde P^n}),
\end{array}
\end{equation}
where $P$ is the pressure, and ${\tilde P}$ the pressure in the absence of
external gravity.

Step 4. Recompute the external potential in the middle time step using the
estimated velocity at the middle time step:
\begin{equation}
	v_i^{n+1/2} = {v_i^n +v_i^{n+1}\over 2}\rightarrow \phi^{n+1/2}.
\end{equation}

Step 5. Finally, we compute the velocity field and gravitational potential
at the $n$+1-th step using the velocity field at the mid step above.
\begin{equation}
	v_i^{n+1} = v_i^{n+1/2}+{\partial \phi^{n+1/2}\over \partial x_i}
			{dt\over 2}.
\end{equation}

In the above procedure, the effect of the black hole's gravity is computed up
to $(\Delta t)^2$. In the vicinity of the black hole, the acceleration is
completely dominated by the black hole and it is important to account for the
black hole gravity very accurately.

\subsection{Conversion from TVD to SPH}
\label{sec:TVD2SPH}

The quantaties of one SPH particle are usually mapped onto several vertices of
the TVD grid.  The number of mapped TVD vertices depends on the size of the
resolution length of the SPH particle.  On the other hand, the inverse mapping
is not mathematically unique since the location and the mass of particles
cannot be simultaneously determined.  It would be possible to create SPH
particles at every TVD vertex on the boundary, but there would be too many SPH
particles.  We have tested possibility of creating
SPH particles by summing up the quantities of $2^3$ or $3^3$ TVD vertices.
By doing this, we were able to decrease the number of created SPH
particles by a large factor, but the gravitational potential energy was not
well conserved because the strong gravitational force by the SMBH made
a small positional error during the summing-up process and
resulted in a quite large potential energy discrepancy.  Thus, we concluded that
the SPH particles have to be created at each TVD vertex.  In order to reduce
the number of SPH particles to a manageable level, we did not create a particle
if the density in a cubical grid is smaller than $10\rho_{min}$, where
$\rho_{min}$ is the density of the ambient medium (i.e., vacuum) in TVD
program.  We used $\rho_{min} \approx 10^{-4} {\bar \rho}$ where $\bar\rho$
is the mean density of the stream.  Also the particle creation was done
only for the case where the mean flux at the cubical grid is directed outward.
Similarly, the velocity and thermal energy of the newly created particles are
determined by the flux and thermal energy density of the corresponding
TVD vertex.

\section{SIMULATION}
\label{sec:simulation}

\subsection{Parameters}

We calculated the evolution of the debris of the tidally disrupted
star by an SMBH using our hybrid code developed in this study.  Only SPH
is used for the calculation until right after the tidal disruption, and the
TVD part starts when the particles with shorter orbital periods return to
the vicinity of the black hole for the second time.  The physical parameters
chosen in this simulation are indentical to those given in \S \ref{sec:sph}.
The natural units used here are $GM_{\rm BH}/c^3 \simeq 49$ sec for the time,
$GM_{\rm BH}/c^2 \simeq 1.48 \times 10^{12}$ cm for the length,
and $M_{\rm BH}$ for the mass.

The size of the TVD grid is $180 \times 250 \times 80$, and the distance
between vertices is 0.5 code length unit.  The TVD box was located such
that the stream-stream collision takes place at the center of the box as
shown in Figure \ref{fig:sph3}.

\subsection{Results}

\begin{figure}
\plotone{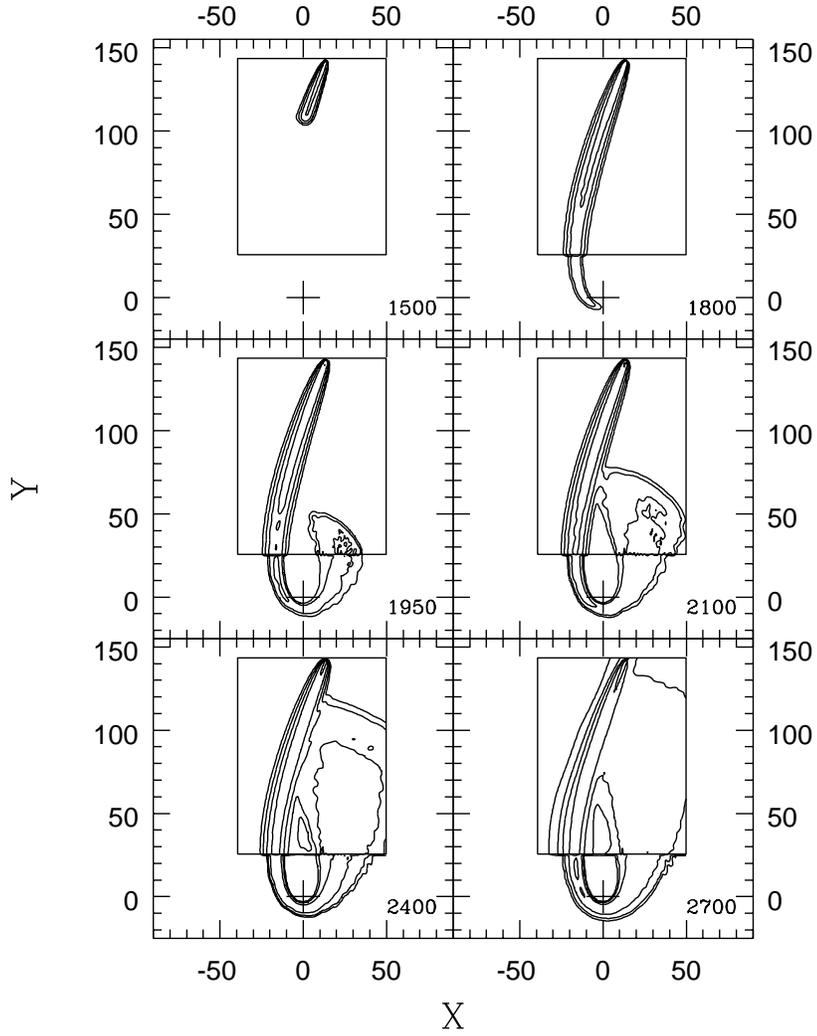}
\caption{\label{fig:contour_snap6}6 density contour maps of the results
of TVD-SPH hybrid simulation for the evolution of tidally disrupted stellar
debris by the SMBH.  The inner boxes represent the TVD box, in which the
caculation is done with the TVD scheme.  The number of TVD grid points for
this simulation is $180\times 250\times 80$, and the length
is in $1.48\times 10^{12}$ cm.  Total duration of this simulation is about
30 hours.}
\end{figure}

Figure \ref{fig:contour_snap6} is a series of density contour snapshots
of our simulation.  The contour lines are drawn at every increment
by a factor of 10.  At $it=1500$, where $it$ is the time step index,
an unperturbed stream enters into the TVD box indicated by the inner box.
At $it=1800$, the stream passes by the SMBH whose location is indicated by
a `+' sign.  The first collision occurs at around $it=2100$.
At $it=2400$, the incoming stream does not seem to be much perturbed by
the collision with the very top of the stream's head.  Instead, the head
of the outgoing stream is reflected by the incoming stream and progresses
in the opposite direction to the incoming stream because the expansion of
the outoing stream during the second revolution around the SMBH makes the
density of it lower than the incoming stream by a factor of a few tens.
This expansion is caused by the finite size of the stream which has been
set somewhat arbitrarily in this study, and will be discussed in more detail
in \S \ref{sec:discussion}.  The effects of the collision on the incoming
stream clearly appear as a perpendicular velocity component in it from
$it=2700$ (see discussions below).  The outer contour lines of the stream
are not smoothly connected at the SPH-TVD boundary at $it=2700$ and 3000,
because the SPH particles are not created for TVD vertices with low
densities, as discussed in \S \ref{sec:TVD2SPH}.  We have stopped our
calculation at $it=3000$.

The velocity field at $it=2700$ is shown in Figure \ref{fig:vector}.
The lengths of the arrows are proportional to the flux-averaged velocity
over the z-axis.  It clearly shows that the low-density outgoing stream
is reflected by the high density barrier of the incoming stream and that
a part of the incoming stream is pushed
away to the opposite side of the collision.  It is interesting to note that
the outgoing tenuous material has nearly uniform velocity field (very
close to the positive y-axis).  This is because the reflected matter exerts
ram pressure to the following material.  If the stream were very thin,
then it would have not expanded after revolving around the SMBH.  The
collision point would be (10,140) (if one neglects the Lense-Thirring
precession).
However, since the stream does have a finite thickness and thus expands after
revolving around the SMBH, the collision is not limited in a small region.
Furthermore, first-colliding matters give following matters pressures toward
positive x-axis direction.  This causes the collision area even
larger and the tail-to-head density ratio at the collision area becomes
bigger.  The conversion efficiency from kinetic energy to thermal energy
becomes weak if the density ratio becomes large.

\begin{figure}
\plotone{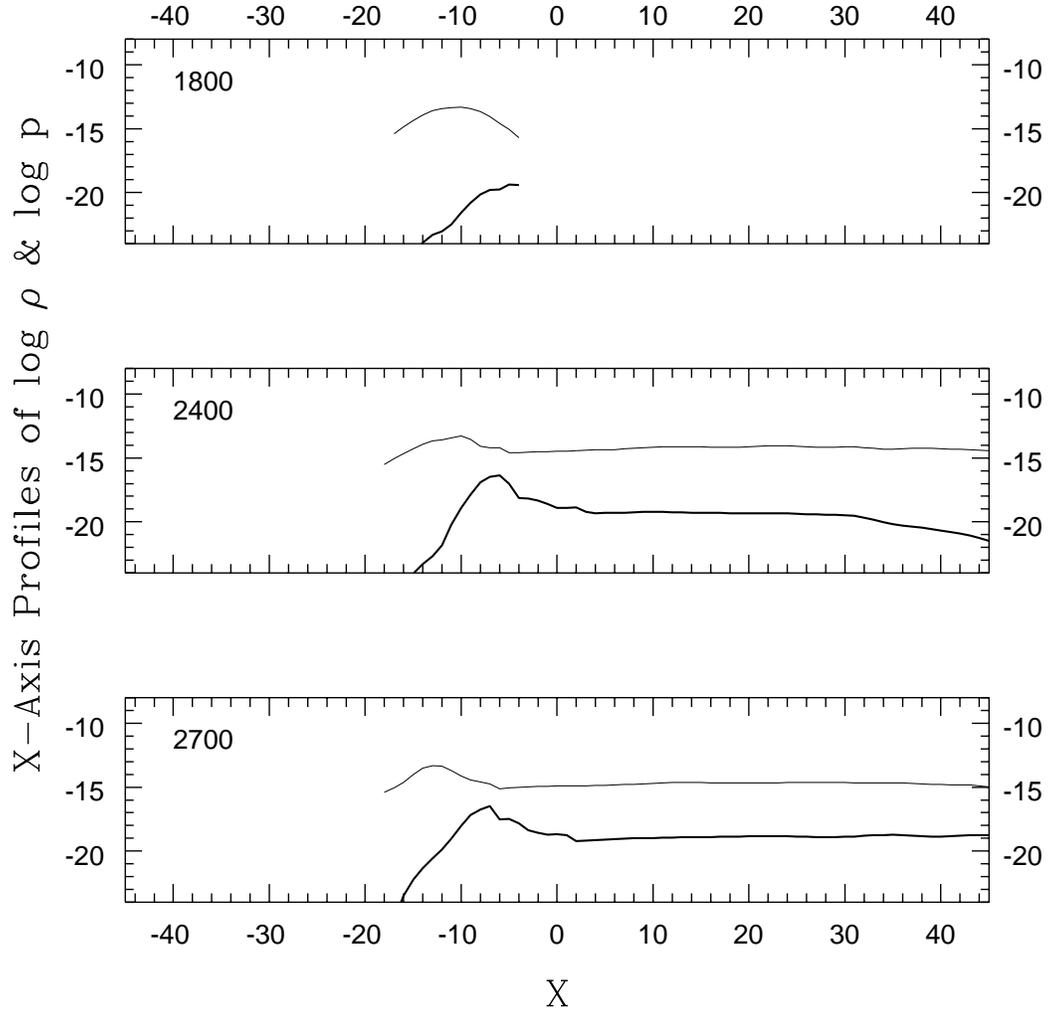}
\caption{\label{fig:cut}X-axis density (thin lines) and pressure (thick lines)
profiles at $Y \simeq 90$ at $it=1800$, 2400, and 2700.  The pressure is
density-averaged over the z-axis.  Densities and pressures are in code units.
(See the text for the code units.)}
\end{figure}

Density and pressure profiles along x-axis at $Y \simeq 90$ for $it=$
1800, 2400, and 2700 are shown in Figure \ref{fig:cut}.
The density is summed over the z-axis and the pressure is flux-averaged
over the z-axis.  At $it=1800$, the density profile is nearly symmetric,
but the pressure profile is inclined to the outside (in view of the SMBH).
This is because the gravitational field of the SMBH is so strong that
a small positional difference makes a non-neglegible velocity
shear.  This effect is stronger at the side closer to the SMBH.
The location of the collision may be found at the pressure peaks at $it=$
2400 and 2700.  Although the tail-to-head density ratio is about $10^2$ at
these steps, the ram pressure of the head is strong enough to push the tail
toward the negative x-axis: between $it=1800$ and $it=2700$, the peak of the
tail's density has been shifted by approximately 2 code length units.
This shift will move the collision point outward when the shifted
matter collides its tail later, making the collision area gradually wider
again.

\section{Discussion}
\label{sec:discussion}

\begin{figure}
\plotone{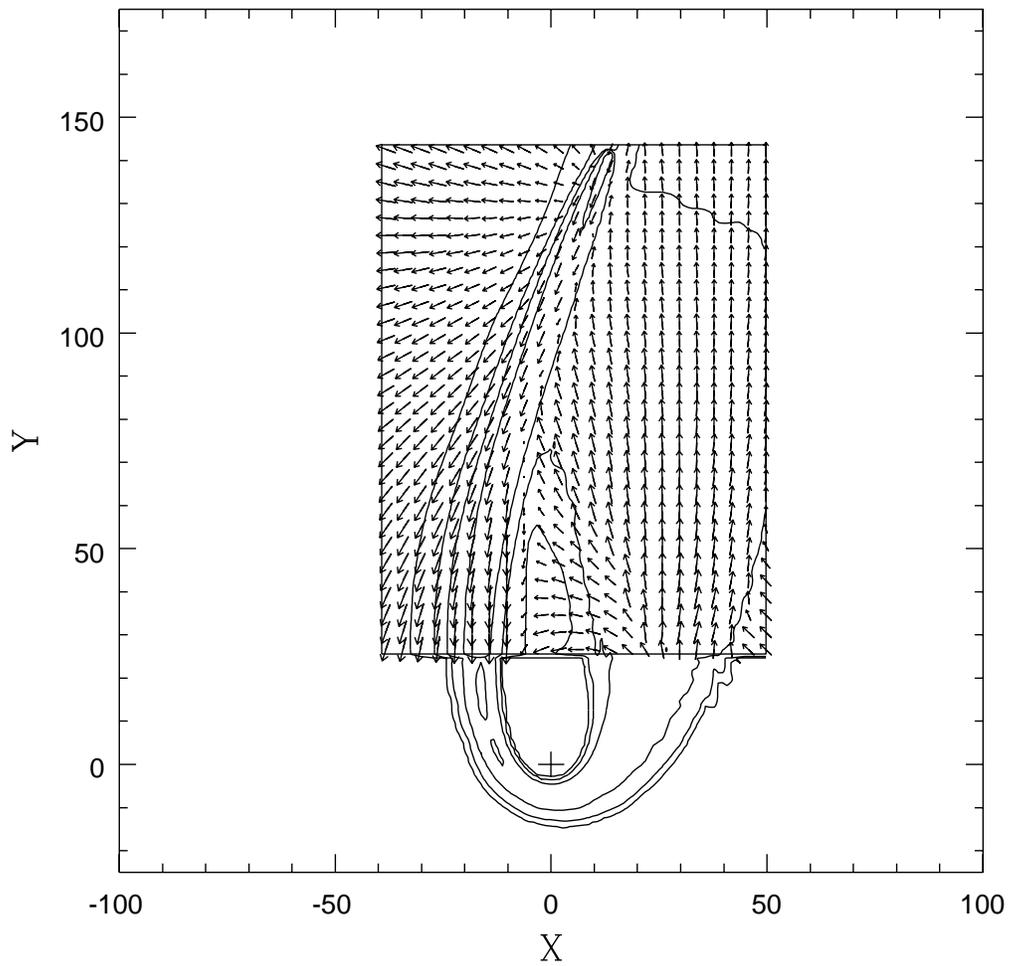}
\caption{\label{fig:vector}Velocity field of our simulation at $it=2700$.
The velocity has been density-averaged over the z-axis.  The result of the
stream collision is more of a reflection than a merger.}
\end{figure}

\begin{figure}
\plotone{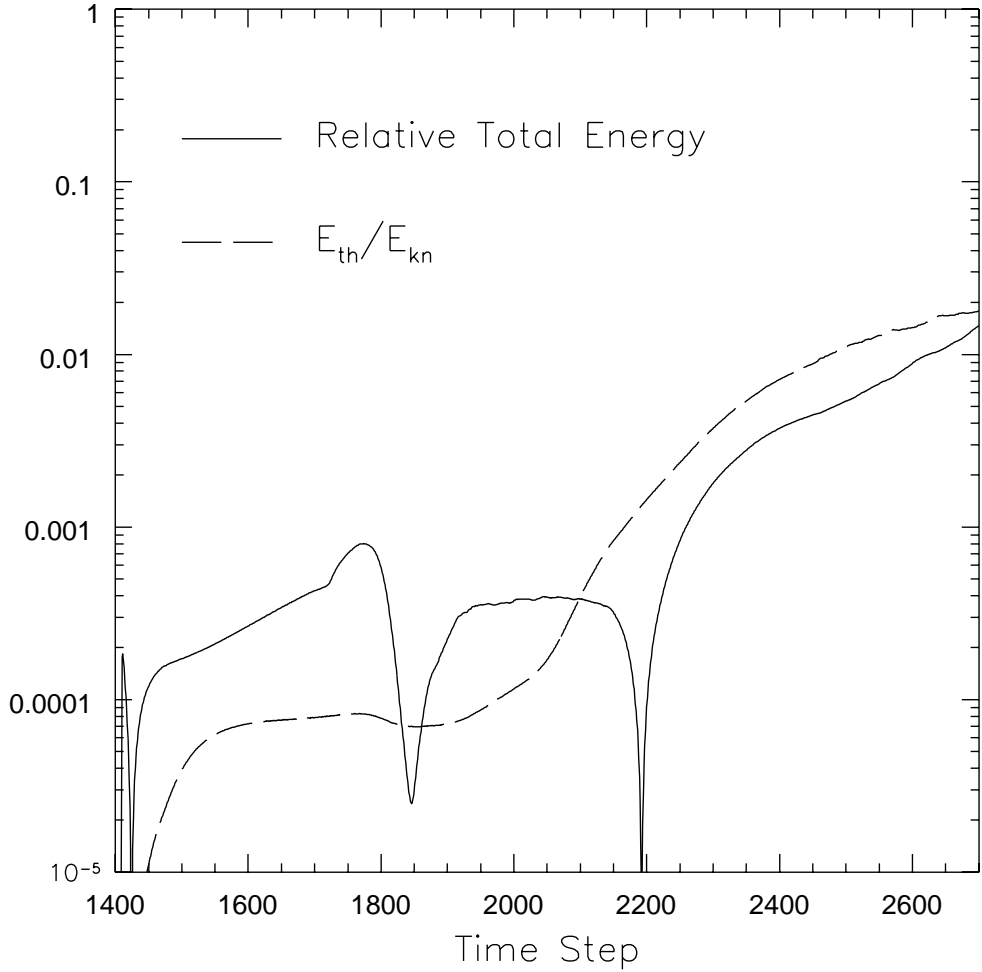}
\caption{\label{fig:energy}Temporal evolution of the relative total energy and
the ratio of thermal energy to kinetic energy.  The relative total energy
is defined as $(E_{gp}+E_{kn}+E_{th})$/$(|E_{gp}|+E_{kn}+E{th})$, where
$E_{gp}$ is the gravitational potential energy, $E_{kn}$ the kinetic energy,
and $E_{th}$ the thermal energy.
The relative total energy is lower than 0.02 for whole simulation, indicating
that the accuracy of the code is acceptable.  The main source of the inaccuracy
is found to come from the coarseness of the TVD grids, not the interfaces
between the SPH and TVD.}
\end{figure}

The major effect of the supersonic stream collision is
transferring the orbital energy into thermal energy which eventually
becomes expansion energy of the shocked region.  Although it has been seen
in Figure \ref{fig:vector} that the ordered velocity component has changed into
somewhat random component, only a small amount of kinetic energy has been
converted into thermal energy ($\sim 2\%$; see Figure \ref{fig:energy}).
We obtain such a low conversin efficiency because the tail-to-head density
ratio at the collision region was as large as 100.  The major effect of such
collision is a reflection of a lower density stream by the high density
stream.

As discussed in the previous section, the expansion of the outgoing stream
in the early stage is due to the finite thickness of the stream: the
thicker the stream, the bigger the expansion.  Kochanek (1994) argued that
there is a focal point near the SMBH where the thickness of the stream becomes
infinitesimal.  If this is real, the density ratio at the collision region
may be close to unity and the collision may be  more effective in
converting kinetic energy into thermal energy.  However, since the resolution
of our TVD grid is of order of the thickness of the stream, such a focal
point could not be realized in our simulation.  To accomplish a resolution
suitable for this phenomenon, we need to increase the number of the TVD grids
by a factor of at least $5^3$, which is of course nearly
impossible with a current CPU and memory limit.  Thus, the significance of
our simulation presented here is rather in making a smoothely working hybrid
code out of two conceptually different numerical methods.

However, the results of our simulation give some insights in predicting the
evolution of the stellar remnant.  Since even with our coarse TVD grids, the
shift of the incoming stream by the ram pressure of the outgoing stream was observed.  This indicates
that the shifted tail will revolve around the SMBH with a slightly different
phase angle and will eventually collide with following stream at farther
distance from the SMBH.  The results of the collision with the stream with a new
phase angle will depend on the amount of the phase angle shift and thickness
of the stream.  Generally, the phase angle change will result in larger
density ratio at the collision area, making the collision less efficient in
converting kinetic energy into thermal energy.  If the phase angle change
induces a change in the density ratio at collision in a much shorter time scale
than the time for returning to the SMBH for the second time of the very tail of
the stream, the collision of the streams will take place intermittently
as suggested by Lee \etal (1996).

The accuracy of our code may be measured with the degree of the
conservation of total energy.  The relative total energy at each time
step is plotted in Figure \ref{fig:energy}.  Until the end of our simulation,
the relative total energy is well conserved by less than 2\%.
The increments of the relative total energy between time steps 1400 and 1700,
and bewteen time steps 2200 and 2700 are caused by the inaccuracy in the TVD
calculation probably due to coarse grids, not by the interfaces between
SPH and TVD.  In fact, the conversions from the TVD to SPH and SPH to TVD
take place after $it=1700$, and the error remains to be very small after
$it=1700$, when the conversion from the TVD to SPH starts, until
the thermal energy created in the coarse TVD box becomes important.

Our calculations assumed only gas pressure.  However, the hot gas in the
shocked region could be dominated by radiation pressure. If we still keep 
the adiabatic assumption but assumes the radiation pressure to dominate,
the adiabatic index would be 4/3. If the radiation energy loss becomes
important, the shock becomes isothermal.  However, the ignorance of
radiative cooling would not be so critical during a very short period
like ours presented here.  Moreover, it is possible that the radiation
loss is not important because of the high opacity for the radiation
behind the shock.  If so, the radiation is simply trapped by the thick medium.

Although it is very difficult to include radiation processes in the simulation
like the one presented here, it is important to account these effects in order
to predict the observational implications.  The stellar disruption could
release significant energy for extended periods through the accretion
onto the black hole.  The stream collision phenomenon could provide a
short-lived activity in galactic nuclei.  All these require more accurate
understanding of stellar disruption.

\acknowledgements

This work was supported by the Cray R\&D Grant in 1995.  H. M. L. was
supported in part by Basic Science Research Institute Program to Pusan
National University under grant No. BSRI 95-2413.


\end{document}